\documentstyle[12pt]{article}

\catcode`\@=11
\long\def\@makefntext#1{
\protect\noindent \hbox to 3.2pt {\hskip-.9pt
$^{{\ninerm\@thefnmark}}$\hfil}#1\hfill}                

 \def\@makefnmark{\hbox to 0pt{$^{\@thefnmark}$\hss}}  

\def\ps@myheadings{\let\@mkboth\@gobbletwo
\def\@oddhead{\hbox{}
\rightmark\hfil\ninerm\thepage}
\def\@oddfoot{}\def\@evenhead{\ninerm\thepage\hfil
\leftmark\hbox{}}\def\@evenfoot{}
\def\sectionmark##1{}\def\subsectionmark##1{}}

\newcounter{sectionc}\newcounter{subsectionc}\newcounter{subsubsectionc}
\renewcommand{\section}[1] {\vspace{0.6cm}\addtocounter{sectionc}{1}
\setcounter{subsectionc}{0}\setcounter{subsubsectionc}{0}\noindent
	{\bf\thesectionc. #1}\par\vspace{0.4cm}}
\renewcommand{\subsection}[1] {\vspace{0.6cm}\addtocounter{subsectionc}{1}
	\setcounter{subsubsectionc}{0}\noindent
	{\it\thesectionc.\thesubsectionc. #1}\par\vspace{0.4cm}}
\renewcommand{\subsubsection}[1]
{\vspace{0.6cm}\addtocounter{subsubsectionc}{1}
	\noindent {\rm\thesectionc.\thesubsectionc.\thesubsubsectionc.
	#1}\par\vspace{0.4cm}}

\newcounter{appendixc}
\newcounter{subappendixc}[appendixc]
\newcounter{subsubappendixc}[subappendixc]

\renewcommand{\appendix}[1] {\vspace{0.6cm}
	\refstepcounter{appendixc}
	\setcounter{figure}{0}
	\setcounter{table}{0}
	\setcounter{equation}{0}
	\renewcommand{\thefigure}{\Alph{appendixc}.\arabic{figure}}
	\renewcommand{\thetable}{\Alph{appendixc}.\arabic{table}}
	\renewcommand{\theappendixc}{\Alph{appendixc}}
	\renewcommand{\theequation}{\Alph{appendixc}.\arabic{equation}}
	\noindent{\bf Appendix \theappendixc #1}\par\vspace{0.4cm}}

\def\abstracts#1{{
	\centering{\begin{minipage}{30pc}\tenrm\baselineskip=12pt\noindent
	\centerline{\tenrm ABSTRACT}\vspace{0.3cm}
	\parindent=0pt #1
	\end{minipage}}\par}}

\renewenvironment{thebibliography}[1]
	{\begin{list}{\arabic{enumi}.}
	{\usecounter{enumi}\setlength{\parsep}{0pt}
\setlength{\leftmargin 1.25cm}{\rightmargin 0pt}
	 \setlength{\itemsep}{0pt} \settowidth
	{\labelwidth}{#1.}\sloppy}}{\end{list}}

\newcounter{itemlistc}
\newcounter{romanlistc}
\newcounter{alphlistc}
\newcounter{arabiclistc}

\newcommand{\fcaption}[1]{
	\refstepcounter{figure}
	\setbox\@tempboxa = \hbox{\tenrm Fig.~\thefigure. #1}
	\ifdim \wd\@tempboxa > 6in
	   {\begin{center}
	\parbox{6in}{\tenrm\baselineskip=12pt Fig.~\thefigure. #1}
	    \end{center}}
	\else
	     {\begin{center}
	     {\tenrm Fig.~\thefigure. #1}
	      \end{center}}
	\fi}

\newcommand{\tcaption}[1]{
	\refstepcounter{table}
	\setbox\@tempboxa = \hbox{\tenrm Table~\thetable. #1}
	\ifdim \wd\@tempboxa > 6in
	   {\begin{center}
	\parbox{6in}{\tenrm\baselineskip=12pt Table~\thetable. #1}
	    \end{center}}
	\else
	     {\begin{center}
	     {\tenrm Table~\thetable. #1}
	      \end{center}}
	\fi}

\def\@citex[#1]#2{\if@filesw\immediate\write\@auxout
	{\string\citation{#2}}\fi
\def\@citea{}\@cite{\@for\@citeb:=#2\do
	{\@citea\def\@citea{,}\@ifundefined
	{b@\@citeb}{{\bf ?}\@warning
	{Citation `\@citeb' on page \thepage \space undefined}}
	{\csname b@\@citeb\endcsname}}}{#1}}

\newif\if@cghi
\def\cite{\@cghitrue\@ifnextchar [{\@tempswatrue
	\@citex}{\@tempswafalse\@citex[]}}
\def\citelow{\@cghifalse\@ifnextchar [{\@tempswatrue
	\@citex}{\@tempswafalse\@citex[]}}
\def\@cite#1#2{{$\null^{#1}$\if@tempswa\typeout
	{IJCGA warning: optional citation argument
	ignored: `#2'} \fi}}

\def\fnt#1#2{\footnotetext{\kern-.3em
	{$^{\mbox{\sevenrm #1}}$}{#2}}}

 1
 1
 1

\font\tenrm=cmr10
\font\tenit=cmti10

\font\ninerm=cmr9

\begin{document}

\begin{flushright} UCRHEP-T137\\November 1994\
\end{flushright}
\vspace{0.8cm}
\centerline{\bf TWO-DOUBLET HIGGS STRUCTURE}
\baselineskip=16pt
\centerline{\bf AT THE ELECTROWEAK ENERGY SCALE\footnote{To appear in the
Proceedings of the 7th Adriatic Meeting on Particle Physics, Brijuni,
Croatia (September 1994)}}
\vspace{0.8cm}
\centerline{\tenrm ERNEST MA}
\baselineskip=13pt
\centerline{\tenit Department of Physics, University of California,}
\baselineskip=12pt
\centerline{\tenit Riverside, California 92521, USA}
\vspace{0.9cm}
\abstracts{The existence of supersymmetry above a few TeV and that of
two Higgs doublets at the electroweak energy scale do not necessarily
result in the minimal supersymmetric standard model (MSSM).  An
interesting counter example is given with $m_h < {\sqrt 2} m_W~+$ radiative
corrections $\sim$ 145 GeV instead of $m_h < m_Z~+$ radiative corrections
$\sim$ 128 GeV in the MSSM.}
\vfil
\rm\baselineskip=14pt
\section{Preamble}
This year, there have been two very reassuring events in high-energy
physics.  One is the possible discovery of the top quark at Fermilab;\cite{1}
and the other is the continuation of this series of Adriatic Meetings on
Particle Physics, now the 7th and for the first time in Brijuni, Croatia.
The Chairman Prof.~Tadic and all the other organizers are to be
congratulated.

\section{Introduction}
The reported central value of the top-quark mass is 174 GeV which is
exactly equal to the vacuum expectation value $v$ of electroweak symmetry
breaking given by
\begin{equation}
{G_F \over \sqrt 2} = {1 \over {4 v^2}}.
\end{equation}
This means that the era of experimental exploration of physics at the
electroweak energy scale has begun.  What is the next particle to be
discovered?  In the standard model, there is just the one Higgs boson,
but in most of its extensions, there are likely to be two Higgs
doublets.\cite{2}  In the very popular minimal supersymmetric standard
model (MSSM), two Higgs doublets are in fact required.  The lightest neutral
scalar boson must also satisfy the mass bound
\begin{equation}
m_h < m_Z + {\rm radiative~corrections},
\end{equation}
which is about 128 GeV for $m_t = 174$ GeV.  This bound is saturated in
the limit of a large pseudoscalar mass (say a few TeV), where the
two-doublet Higgs structure of the MSSM reduces to only one Higgs boson
at the electroweak energy scale as in the standard model.

The implicit assumption of the MSSM is that at the energy scale of soft
supersymmetry breaking, say a few TeV, the gauge group is the standard
$\rm SU(3) \times SU(2) \times U(1)$.  If the latter is something larger
but it breaks down to the standard gauge group also at a few TeV, then
the structure of the Higgs potential is determined by the scalar particle
content needed for that symmetry breaking.  Furthermore, the quartic
scalar couplings are related to the gauge couplings of the larger theory
as well as other couplings appearing in its superpotential.
\vspace{0.2cm}

At the electroweak energy scale, the reduced Higgs potential may contain only
two scalar doublets, but their quartic couplings are generally not those
of the MSSM.  In this talk I will describe one such counter example\cite{3}
based on a very interesting supersymmetric left-right model proposed some
years ago.\cite{4}  In my second talk\cite{5} I will give more details of
that model itself and describe some recent results on its possible
unification at the $10^{16}$ GeV energy scale, as well as its effect on
the precision measurements of $Z \rightarrow$ leptons.

\section{The Two-Doublet Higgs Potential}
Consider two Higgs doublets $\Phi_{1,2} = (\phi^+_{1,2}, \phi^0_{1,2})$
and the Higgs potential
\begin{eqnarray}
V &=& \mu_1^2 \Phi_1^\dagger \Phi_1 + \mu_2^2 \Phi_2^\dagger \Phi_2 +
\mu_{12}^2 (\Phi_1^\dagger \Phi_2 + \Phi_2^\dagger \Phi_1) \nonumber \\
&+& {1 \over 2} \lambda_1 (\Phi_1^\dagger \Phi_1)^2 + {1 \over 2} \lambda_2
(\Phi_2^\dagger \Phi_2)^2 + \lambda_3 (\Phi_1^\dagger \Phi_1) (\Phi_2^\dagger
\Phi_2) \nonumber \\ &+& \lambda_4 (\Phi_1^\dagger \Phi_2) (\Phi_2^\dagger
\Phi_1) + {1 \over 2} \lambda_5 (\Phi_1^\dagger \Phi_2)^2 + {1 \over 2}
\lambda_5^* (\Phi_2^\dagger \Phi_1)^2.
\end{eqnarray}
In the MSSM, there are the well-known constraints
\begin{equation}
\lambda_1 = \lambda_2 = {1 \over 4} (g_1^2 + g_2^2), ~~~ \lambda_3 =
- {1 \over 4} g_1^2 + {1 \over 4} g_2^2, ~~~ \lambda_4 = - {1 \over 2} g_2^2,
{}~~~ \lambda_5 = 0,
\end{equation}
where $g_1$ and $g_2$ are the U(1) and SU(2) gauge couplings of the standard
model respectively.  As $\phi_{1,2}^0$ acquire vacuum expectation values
$v_{1,2}$, two tree-level sum rules are obtained:
\begin{equation}
m_{h^0}^2 + m_{H^0}^2 = m_Z^2 + m_A^2, ~~~ m_{H^\pm}^2 = m_W^2 + m_A^2,
\end{equation}
where the pseudoscalar mass $m_A$ is given by
\begin{equation}
m_A^2 = - \mu_{12}^2 (\tan \beta + \cot \beta), ~~~ \tan \beta \equiv v_2/v_1.
\end{equation}
Note that only the gauge couplings contribute to the $\lambda$'s.  This is
because that with only two $\rm SU(2) \times U(1)$ Higgs superfields,
there is no cubic invariant in the superpotential and thus no additional
coupling.

\section{The E$_6$-Inspired Left-Right Model}
Consider now the gauge group $\rm SU(2)_L \times SU(2)_R \times U(1)$
but with an unconventional assignment of fermions\cite{4} including an
exotic quark $h$ of electric charge $-1/3$.
\begin{eqnarray}
\left(  \! \begin{array} {c} u \\ d \end{array} \! \right)_L \sim (2,1,1/6),
{}~&~&~ d_R \sim (1,1,-1/3), \\ \left( \! \begin{array} {c} u \\ h \end{array}
\! \right)_R \sim (1,2,1/6), ~&~&~ h_L \sim (1,1,-1/3), \\ \Phi_1 \equiv
\left( \! \begin{array} {c} \phi_1^+ \\ \phi_1^0 \end{array} \! \right) \sim
(2,1,1/2), ~&~&~ \chi \equiv \left( \! \begin{array} {c} \chi^+ \\ \chi^0
\end{array} \! \right) \sim (1,2,1/2), \\ \eta \equiv \left( \begin{array}
{c@{\quad}c} \overline {\phi_2^0} & \eta^+ \\
- \phi_2^- & \eta^0 \end{array} \right) &\sim& (2,2,0).
\end{eqnarray}
Note that the mass matrices for the $u$, $d$, and $h$ quarks are proportional
to different vacuum expectation values, {\it i.e.} $\langle \phi_2^0 \rangle$,
$\langle \phi_1^0 \rangle$, and $\langle \chi^0 \rangle$ respectively.
Hence flavor-changing neutral
currents are guaranteed to be absent at tree level in this model,
as opposed to the conventional left-right model where they are
unavoidable.  Note also that $\Phi_1^\dagger \tilde \eta \chi$ is now
an allowed term in the superpotential, where $\tilde \eta \equiv \sigma_2
\eta^* \sigma_2$, so that its coupling $f$ also contributes to the quartic
scalar couplings of this model's Higgs potential.

Let $G_1$ be the U(1) gauge coupling and $G_2$ the coupling of both SU(2)'s.
Then
\begin{eqnarray}
V &=& V_{soft} + {1 \over 8} (G_1^2 + G_2^2)[(\Phi_1^\dagger \Phi_1)^2 +
(\chi^\dagger \chi)^2] \nonumber \\ &+& {1 \over 4} G_2^2 [({\rm Tr}
\eta^\dagger \eta)^2 - ({\rm Tr} \eta^\dagger \tilde \eta) ({\rm Tr} \tilde
\eta^\dagger \eta)] + (f^2 - {1 \over 4} G_2^2)(\Phi_1^\dagger \Phi_1 +
\chi^\dagger \chi) {\rm Tr} \eta^\dagger \eta \nonumber \\ &-& (f^2 -
{1 \over 2}G_2^2)(\Phi_1^\dagger \eta \eta^\dagger \Phi_1 + \chi^\dagger
\eta^\dagger \eta \chi) + (f^2 - {1 \over 4} G_1^2)(\Phi_1^\dagger \Phi_1)
(\chi^\dagger \chi),
\end{eqnarray}
where $V_{soft}$ contains terms of dimensions 2 and 3, and breaks the
supersymmetry.  Let $\chi^0$ acquire a vacuum expectation value $u \neq 0$.
Then $\rm SU(2)_L \times SU(2)_R \times U(1)$ breaks down to the standard
$\rm SU(2)_L \times U(1)_Y$ with $m^2(\sqrt 2 Re \chi^0) = (G_1^2 + G_2^2)
u^2/2$ and $m^2(\eta^+,\eta^0) = G_2^2 u^2/2$.  These heavy particles can be
integrated out at the electroweak energy scale where only $\Phi_{1,2}$ are
left.

\section{Reduced Higgs Potential of the Left-Right Model}
The quartic scalar couplings of the reduced Higgs potential at the
electroweak energy scale are now given by
\begin{eqnarray}
\lambda_1 &=& {1 \over 4} (G_1^2 + G_2^2) - {{(4 f^2 - G_1^2)^2} \over
{4 (G_1^2 + G_2^2)}}, \\ \lambda_2 &=& {1 \over 2} G_2^2 - {{(4 f^2 -
G_2^2)^2} \over {4 (G_1^2 + G_2^2)}}, \\ \lambda_3 &=& {1 \over 4} G_2^2
- {{(4 f^2 - G_1^2)(4 f^2 - G_2^2)} \over {4 (G_1^2 + G_2^2)}}, \\
\lambda_4 &=& f^2 - {1 \over 2} G_2^2, ~~~ \lambda_5 ~=~ 0,
\end{eqnarray}
where the second terms on the right-hand sides of the equations for
$\lambda_{1,2,3}$ come from the cubic interactions of $\sqrt 2 Re \chi^0$
which are proportional to $u$.  When divided by the square of its mass, these
contributions do not vanish even if $u$ becomes very large.  This is
another example of nondecoupling.

In the limit $f = 0$ and using the tree-level boundary conditions
\begin{equation}
G_2 = g_2, ~~~ G_1^{-2} + G_2^{-2} = g_1^{-2},
\end{equation}
it can easily be shown from the above that the MSSM is recovered.
However, $f$ is in general nonzero, although it does have an upper bound
because $V$ must be bounded from below.  Hence
\begin{equation}
0 \leq f^2 \leq {1 \over 4} (g_1^2 + g_2^2) \left( 1 - {g_1^2 \over g_2^2}
\right)^{-1},
\end{equation}
where the maximum value is obtained if $V_{soft}$ is aslo left-right
symmetric.  For illustration, let $f = f_{max}$ and $x \equiv \sin^2
\theta_W$, then
\begin{equation}
\lambda_1 = 0, ~~~ \lambda_2 = {e^2 \over {2x}} \left[ 1 - {{2 x^2} \over
{(1-x)(1-2x)}} \right],
\end{equation}
\begin{equation}
\lambda_3 = {e^2 \over {4x}} \left[ 1 - {{2x} \over {1-2x}} \right] =
- \lambda_4, ~~~ \lambda_5 = 0.
\end{equation}
It is clear that the MSSM conditions of Eq. (4) are no longer appropriate.

\section{Phenomenological Consequences}
At tree level, the $2 \times 2$ mass-squared matrix of the two neutral
Higgs scalar bosons of the two-doublet model is given by
\begin{equation}
{\cal M}^2_{h,H} = \left[ \begin{array} {c@{\quad}c} 2 \lambda_1 v_1^2 -
\mu_{12}^2 v_2/v_1 & 2 (\lambda_3 + \lambda_4) v_1 v_2 + \mu_{12}^2 \\
2 (\lambda_3 + \lambda_4) v_1 v_2 + \mu_{12}^2 & 2 \lambda_2 v_2^2 -
\mu_{12}^2 v_1/v_2 \end{array} \right].
\end{equation}
Hence there is an upper bound on the lighter of the two bosons:
\begin{equation}
m_h^2 < 2(v_1^2 + v_2^2) [\lambda_1 \cos^4 \beta + \lambda_2 \sin^4 \beta
+ 2(\lambda_3 + \lambda_4) \sin^2 \beta \cos^2 \beta] + \epsilon,
\end{equation}
where $\tan \beta \equiv v_2/v_1$ as defined previously, and $\epsilon$
is a radiative correction due to the presence of the top quark:
\begin{equation}
\epsilon = {{3 g_2^2 m_t^4} \over {8 \pi^2 m_W^2}} \ln \left( 1 +
{\tilde m^2 \over m_t^2} \right),
\end{equation}
where $\tilde m$ is an effective mass of the two scalar top quarks. The
MSSM bound is given by
\begin{equation}
m_h^2 (f = 0) < (m_Z \cos 2 \beta)^2 + \epsilon,
\end{equation}
whereas in this model with $f = f_{max}$, it is given by
\begin{equation}
m_h^2 (f = f_{max}) < 2 m_W^2 \left[ 1 - {{2 \sin^4 \theta_W} \over
{\cos^2 \theta_W \cos 2 \theta_W}} \right] \sin^4 \beta + \epsilon.
\end{equation}
However, $m_h$ is maximized not at $f_{max}$, but at\cite{6}
\begin{equation}
f_0^2 = {1 \over 4} g_2^2 \left( 1 - \cos^4 \beta + {g_1^2 \over g_2^2}
\cos 2 \beta \right) \left( 1 - {g_1^2 \over g_2^2} \right)^{-1}
\end{equation}
which is less than $f_{max}^2$ for all $\beta$.  The upper bounds on $m_h$
for $f = f_0$, $f = f_{max}$, and $f = 0$ are plotted together in Fig. 1.
The absolute upper bound on $m_h$ is then given by
\begin{equation}
m_h^2 < 2 m_W^2 + \epsilon,
\end{equation}
which occurs at $\cos^2 \beta = 0$.  Using $m_t = 174$ GeV and $\tilde m = 1$
TeV, this implies that
\begin{equation}
m_h < 145~{\rm GeV}
\end{equation}
in this model, as opposed to the upper bound of 128 GeV in the MSSM ($f=0$).

\section{Another Scenario}
It has been shown in the above that the existence of a cubic term in the
superpotential will change the two-doublet Higgs structure at the
electroweak energy scale.  This occurs naturally for a gauge symmetry
larger than the standard $\rm SU(2) \times U(1)$ because the superfields
involved are likely to be precisely those required for the symmetry
breaking and the generation of fermion masses, as is the case of the
E$_6$-inspired supersymmetric left-right model.\cite{4}  However, even
if the gauge symmetry is only $\rm SU(2) \times U(1)$, a singlet scalar
superfield $N$ may be added so that the term $f \Phi_1^\dagger \Phi_2 N$
is allowed.  Now $\lambda_{1,2,3,5}$ are as in the MSSM, but
\begin{equation}
\lambda_4 = - {1 \over 2} g_2^2 + f^2.
\end{equation}
Assume that $\langle N \rangle = 0$, then
\begin{equation}
m^2_{H^\pm} = m_W^2 + m_A^2 - f^2 (v_1^2 + v_2^2),
\end{equation}
hence $m_{H^\pm} < m_W$ is now possible, in contrast to the MSSM or the
left-right model discussed above.  Note that $\lambda_4$ gets the extra
contribution $f^2$, regardless of how heavy $N$ is.  This shows how
sensitive the supersymmetric two-doublet Higgs structure is to the
details of a possible larger theory.
\vspace{0.2cm}

Although the other sum rule of the MSSM [the first one in Eq. (5)] still
holds in this scenario, the upper bound on $m_h$ is different because of
$f$.  For a given value of $m_A$, this upper bound is maximized at
$f^2 (v_1^2 + v_2^2) = (m_A^2 + m_Z^2)/2$, making ${\cal M}^2_{h,H}$ of
Eq. (18) diagonal.  Hence
\begin{equation}
m_h^2 < m_Z^2 \cos^2 \beta + m_A^2 \sin^2 \beta,
\end{equation}
or
\begin{equation}
m_h^2 < m_Z^2 \sin^2 \beta + m_A^2 \cos^2 \beta + {\epsilon \over {\sin^2
\beta}},
\end{equation}
whichever is smaller.  Since the value of $m_A$ is unrestricted, this means
that there is no real upper bound on $m_h$.  The lesson to be learned here
is that with a minimal change in the larger theory, the two-doublet Higgs
structure at the electroweak energy scale can be drastically different.
In the above case of the addition of a seemingly harmless singlet, both
of the well-known bounds of the MSSM on the masses of scalar bosons are
removed.

\section{Conclusions}
(1) Even if supersymmetry exists and there are only two Higgs doublets
at the electroweak energy scale, the minimal supersymmetric standard model
(MSSM) is not the only possibility.
\vspace{0.2cm}

(2) For $m_t = 174$ GeV, $m_h > 128$ GeV rules out the MSSM, and $m_h > 145$
GeV rules out the E$_6$-inspired supersymmetric left-right model.
\vspace{0.2cm}

(3) If two Higgs doublets are discovered and their masses determined,
it may even be possible to deduce what precisely the larger theory is
at the scale of supersymmetry breaking.

\section{Acknowledgements}
I thank Profs. D. Tadic and I. Picek and the other organizers of the 7th
Adriatic Meeting on Particle Physics for their great hospitality and a
very stimulating program.  This work was supported in part by the U. S.
Department of Energy under Grant No. DE-FG03-94ER40837.

\end{document}